\documentclass[aps,preprint,prd,showpacs,nofootinbib]{revtex4}

\usepackage{latexsym}
\usepackage{amsmath}
\usepackage{graphicx}
\usepackage{subfigure}
\usepackage{dcolumn}
\usepackage{bm}
\usepackage{amssymb}
\usepackage{latexsym}
\usepackage{ulem}
\usepackage{color}
\usepackage[colorlinks,linkcolor=magenta,anchorcolor=cyan,citecolor=blue]{hyperref}

\def\be{\begin{equation}}
\def\ee{\end{equation}}
\def\ba{\begin{eqnarray}}
\def\ea{\end{eqnarray}}

\def\lf{\left}
\def\rt{\right}

\begin{document}

\title{Primordial gravastar from inflation}

\author{Yu-Tong Wang$^{1}$\footnote{wangyutong@ucas.ac.cn}}
\author{Jun Zhang$^{2}$$\footnote{jun34@yorku.ca}$}
\author{Yun-Song Piao$^{1,3}$\footnote{yspiao@ucas.ac.cn}}

\affiliation{$^1$ School of Physics, University of Chinese Academy
of Sciences, Beijing 100049, China}

\affiliation{$^2$ Department of Physics and Astronomy, York University, Toronto, Ontario, M3J 1P3, Canada}

\affiliation{$^3$ Institute of Theoretical Physics, Chinese
Academy of Sciences, P.O. Box 2735, Beijing 100190, China}

\begin{abstract}

The dS bubbles can nucleate spontaneously during inflation, and
will be stretched by the cosmological expansion to astrophysical
scale. We report on a novel phenomenon that such a bubble might
develop into a gravastar (an ultra-compact object with a dS core)
after inflation, which witnessed the occurrence of inflation and
would survive till today. It is pointed out that if a primordial
gravastar was involved in one of the LIGO/Virgo gravitational wave
(GW) events, the post-merger object could be a gravastar that will
eventually collapse into a black hole. As a result, the late-time
GW ringdown waveform will exhibit a series of ``echoes" with
intervals increasing with time.

\end{abstract}

\maketitle


The recent direct detections of gravitational waves (GWs) made by the LIGO Scientific and the Virgo Collaborations have opened a new window to probe the strong gravity physics. The detections are usually attributed to coalescences of binary black holes (BHs)
\cite{Abbott:2016blz} or neutron stars \cite{TheLIGOScientific:2017qsa}. However, it is still significant to ask if they could be explained by coalescences of other ultra-compact objects \cite{Cardoso:2017cqb}. Particularly, it has been showed in Refs.\cite{Cardoso:2016rao}\cite{Cardoso:2016oxy} that if the post-merger object is a horizonless exotic compact
object, such as a gravitational vacuum star (GVS) called gravastar
\cite{Mazur:2001fv}\cite{Visser:2003ge}, the GW ringdown waveform
will consist of the primary signal, which is almost identical to
that of a ringing-down BH, and a series of ``echoes". As the echoes encode
informations of new physics or quantum gravity \cite{Giddings:2017jts}, the corresponding signals have been searched in the GW data by many research groups \cite{Abedi:2016hgu}\cite{Abedi:2017isz}\cite{Westerweck:2017hus}\cite{Conklin:2017lwb}.

Actually, like the primordial BHs
\cite{Carr:1974nx}\cite{Carr:1975qj}\cite{Bird:2016dcv}\cite{Clesse:2016vqa}\cite{Sasaki:2016jop},
the GVSs might also be primordial, and could play distinct roles
in cosmological evolution. It has been pointed out in
Ref.\cite{Raidal:2018eoo} that the light primordial GVSs could be
a perfect candidate for cold dark matter (CDM). Moreover, the
interaction between the GVS DM with baryons can help to cool the
baryons at the cosmic dawn, and so explain the EDGES's recent
observation \cite{Bowman:2018yin}\cite{Barkana:2018lgd}. However,
if they exist, how can the primordial GVSs come into being? It is
still an open question.


The spacetime of a GVS can be described by the Schwarzschild
metric, except the Schwarzschild horizon and the region inside are
replaced by a dS core \cite{Mazur:2001fv}. Nonetheless, the
physical origin of the dS core is unclear, see
e.g.\cite{Cardoso:2017cqb}. On the other hand, the dS bubbles can
spontaneously nucleate during inflation \cite{Guth:1982pn}, and be
stretched by the cosmological expansion to astrophysical scales
\cite{Basu:1991ig}\cite{Garriga:2015fdk}. A natural idea would be
that the precursor of a GVS might be just such a dS bubble. In
this report, we will explore the possibility that a primordial
GVS-like object developed from a dS bubble that nucleated during
inflation. We will also discuss its implications for the
LIGO/Virgo events.


Generally, inflation is driven by a scalar field slow rolling
along its effective potential $V_{inf}$. In a Landscape
\cite{Kachru:2003aw}\cite{Susskind:2003kw}, nucleations of bubbles with low-energy ($V_{b}<V_{inf}$) vacua occur inevitably. During inflation, the bubble radius $r_*$ will be stretched
by the expansion of universe, so that after the inflation ends,
$r_*$ may be far larger than $1/H_{inf}$. Dependent on the
nucleation rate $\lambda$ per Hubble volume ($=1/H_{inf}^4$), the
number density of bubble $dn=\lambda dr_*/(r_*+H_{inf}^{-1})^4$ can
have a special distribution \cite{Basu:1991ig}.

When the inflation ends, the inflaton energy $0<\rho_{inf}\simeq
V_{inf}$ outside bubbles will release into radiation. The bubble
wall moves highly relativistic related to matter, but will become
at rest with respect to matter in a very short
time\cite{Garriga:2015fdk}. Once the bubble wall is at rest with
respect to the Hubble flow, it will be surrounded by an
infinitesimal layer devoid of matter with a negligible vacuum
energy, see also \cite{Deng:2017uwc}. In this case, the metric is
\begin{align}
ds^{2}=-f(r) dt^{2}+{dr^{2}\over f(r)}+r^{2}d\Omega^{2},
\end{align}
with
\ba f(r)= \left\{
                        \begin{array}{ll}  1-{2{\cal M}\over r} \quad\, {\rm for} \quad r>r_*, \\
                                    1-H_b^2 r^2 \quad\, {\rm for} \quad r<r_*,
                \end{array}
                \right.
\ea
which is Schwarzschild outside the bubble wall ($r>r_*$), and is de Sitter inside the bubble ($r<r_*$). We assumed the bubble wall is located at $r=r_*$, $H_b^2=8\pi V_b/3$ and
$G=1$.

\begin{figure}[htbp]
\includegraphics[width=0.40\textwidth]{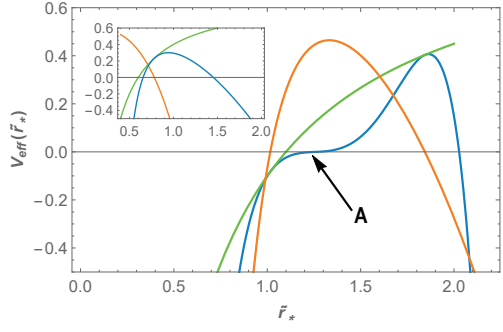}
\caption{The effective potential $V_{eff}({\tilde r}_*)$ (blue
curve), $\beta_S$ (orange curve) and $1-2{\tilde {\cal M}}/{\tilde
r_*}$ (green curve) with respect to ${\tilde r_*}$. The inset is
plotted for $V_{eff}({\tilde r}_*)$, $\beta_S$ and $1-2{\tilde
{\cal M}}/{\tilde r_*}$ but with $\sigma=const$, which is
consistent with Fig.6 in Ref.\cite{Blau:1986cw}. } \label{fig01}
\end{figure}

The equation of motion for the bubble wall $r_*$ is given by
\cite{Blau:1986cw} \be \beta_D -\beta_S=H_\sigma r_*, \ee with \be
|\beta_D|=\lf(1+{\dot
r}_*^2-H_b^2r_*^2\rt)^{1/2},\,\,\,\,\,\,\,|\beta_S|=\lf(1+{\dot
r}_*^2-{2{\cal M}\over r_*}\rt)^{1/2}, \label{beta} \ee which is
equivalent to \be {\dot r}_*^2 +V(r_*)=0, \ee where \be V({
r}_*)=1-(1-{H_b^2\over H_\sigma^2}){{\cal M}\over r_*}-{{\cal
M}^2\over H_\sigma^2r_*^4}-\lf(H_\sigma^2/4+{H_b^2/2}+{H_b^4\over
4H^2_\sigma}\rt)r_*^2, \label{V1}\ee $H_\sigma=4\pi {\sigma}$ with
$\sigma$ being the bubble wall tension, and ${\dot r}={dr/d\tau}$
with $\tau$ being the proper time of the bubble wall. It has been
showed in Ref.\cite{Blau:1986cw} that if $\sigma$ is constant,
$r_*$ will either decrease to $0$ after reaching its maximal
value, or simply grow without bound, see the insert in
Fig.\ref{fig01}. As a result, a bubble will either stop expanding
and collapse to a Schwarzschild singularity, or it will develop
into a baby universe that initially connects to the parent
universe (the exterior of the bubble) through a wormhole and will
detach as the wormhole pinches off eventually. In both cases, the
primordial BHs will form \cite{Garriga:2015fdk}, see also
\cite{Sato:1981bf}\cite{Berezin:1982ur}.

Generally, if the bubble wall consists of the fluids with a state
equation $w$, one will have $\sigma={\sigma_0/r_*^{2(1+w)}}$ with
$\sigma_0$ being a constant, see
e.g.\cite{Chen:2015lbp}\cite{Danielsson:2017riq} for relevant
physics. The scalar field bubble wall corresponds to the fluids
with $w=-1$, and hence $\sigma=\sigma_0$. In
Ref.\cite{Mazur:2001fv}, the shin shell (or wall) with the stiff
fluid ($w=1$) has been adopted for GVS. In the following, we will
consider such a scenario that after inflation the bubble wall will
go through a rapid phase transition (PT) at $t_0$, and after the
PT, the bubble wall will consist of the stiff fluid, so that the
wall tension $\sigma=\sigma_0(r_0/r_*)^4$, where $r_0$ is the
radius of bubble wall at $t\simeq t_0$. Defining $x=4\pi \sigma_0
{\tilde r}_0^4$, we have \be \lf({d {\tilde r}_*/x\over
d\tau}\rt)^2+V_{eff}({\tilde r}_*)=0, \ee where \be
V_{eff}({\tilde r}_*)=1-\lf(1-{\tilde H}_b^2{\tilde
r}_*^8\rt){{\tilde{\cal M}}\over {\tilde r}_*}-{\tilde {\cal
M}}^2{\tilde r}_*^4-\lf({1\over 4{\tilde r}_*^8}+{{\tilde
H}_b^2\over2}+{{\tilde H}_b^2\over 4}{\tilde r}_*^{8}\rt){\tilde
r}_*^2 \label{V2} \ee with the dimensionless parameters ${\tilde
{\cal M}}={x{\cal M}}$, ${\tilde r}_*={r_*x}$ and ${\tilde
H}_b=H_b /x$. Different from the case of constant $\sigma$,
depending on the values of ${\tilde {\cal M}}$ and ${\tilde
H}_b^2$, we may have a metastable static solution for the bubble
wall. Numerically solving Eqs. $V_{eff}({\tilde r}_*)=0$,
$\partial V_{eff}/\partial {\tilde r}_*= 0$ and $\partial^2
V_{eff}/\partial {\tilde r}_*^2= 0$, we find ${\tilde{\cal
M}}=0.55$, ${\tilde H}_b=0.4$ and ${\tilde r}_*\simeq 1.2$, which
correspond to the static point `A' shown in Fig.\ref{fig01}. When
the bubble wall expands and arrives at the `A' point, it will be
at rest for a long time. In this case, the asymptotic radius is
\be {r_*\over {\cal M}}={{\tilde r}_*\over {\tilde {\cal M}}}\sim
2.18, \ee which implies the bubble is slightly larger than the
Schwarzschild radius, and lies inside the light-ring (or
photosphere) of the corresponding Schwarzschild mass ($r=3{\cal
M}$). Also, we have the bubble radius $r_*=({\tilde r}_*{\tilde
H}_b)/H_b\simeq 0.5/H_b<1/H_b$. Therefore, a primordial GVS-like
object comes into being. The existence of the light-ring indicates
that the GVS looks like a BH to a distance observer, and belongs
to the ultra-compact objects \cite{Cardoso:2017njb}. The conformal
diagram of the primordial GVS developing from a dS bubble is shown
in Fig.~2. ($\beta_S={\tilde{\cal M}}{\tilde r}_*^2-{1/(2{\tilde
r}_*^3)}-{\tilde H}_b^2{\tilde r}_*^5/2>0$ suggests that the polar
angle in Schwarzschild space coordinates is increasing
\cite{Basu:1991ig}.)


\begin{figure}[htbp]
\includegraphics[width=0.5\textwidth]{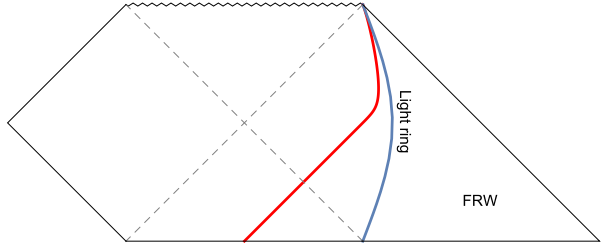}
\caption{The conformal diagram of the spacetime outside the
bubble. The Schwarzschild metric has been matched on a
matter-dominated FRW universe. The red line shows the trajectory
of the bubble wall, which expands initially and will approach to a
constant after the wall went through a PT . The blue line shows
the location of the light-ring (the photosphere) of the
corresponding Schwarzschild mass. Notice that the interior of the
bubble (region on the left hand side of the red line) should be
replaced by the dS spacetime.} \label{fig02}
\end{figure}


The mass ${\cal M}_{\rm PGVS}={{\tilde {\cal M}}/(x)}$ of the
primordial GVS is \be {\cal M}_{\rm PGVS}={{\tilde {\cal
M}}{\tilde H}_b\over H_b}\sim {M_P^2\over H_b}\simeq \lf({1{\rm
Gev}\over \Lambda_b}\rt)^2M_\odot,\label{M} \ee where $H_b\simeq
\sqrt{V_b}/M_P= \Lambda_b^2/M_P$ is used. This indicates that for
$\Lambda_b \sim 1$Gev, we will have the primordial GVS with ${\cal
M}_{\rm PGVS}\sim 1M_\odot$ (the radius $r_{PGVS}\sim
10^3m<1/H_b$). However, ${\cal M}_{\rm PGVS}$ may also be
estimated straightly by using Eq.(\ref{beta}), \be {\cal
M}=\lf({H_b^2-H_\sigma^2\over 2}\rt){r_*^3}+H_\sigma
r_*^2\sqrt{1+{\dot r}_*^2-H_b^2r_*^2}. \ee When the bubble wall
arrives at the `A' point, ${\dot r}_{*A}=0$, so for
$H_{\sigma_A}\lesssim H_b$ and $r_{*A}\simeq 0.5/H_b$, we will get
Eq.(\ref{M}) approximately. Interestingly, ${\cal M}_{\rm PGVS}$
given by Eq.(\ref{M}) actually corresponds to the critical mass
defined in Ref.\cite{Garriga:2015fdk}.

\begin{figure}[htbp]
\includegraphics[width=0.45\textwidth]{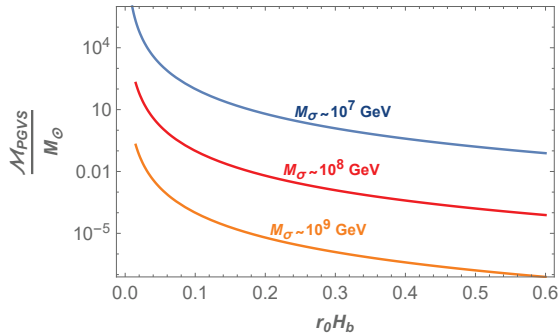}
\caption{The mass ${\cal M}_{PGVS}$ of primordial GVS with respect
to $M_\sigma=\sigma^{1/3}$ and $r_0/(1/H_b)$. } \label{fig03}
\end{figure}

In this scenario,
for the different values of $\sigma_0$, the PT of bubble wall must
occur at the different time $t_0$, and so does the bubble radius
$r_0$. Combing ${\tilde H}_b=H_b /x$, ${\tilde r}_*={r_*x}$ and
$x=4\pi \sigma_0 {\tilde r}_0^4$, we have \be r_0\sim
\lf(M_P^2\over H_b^3\sigma_0\rt)^{1/4}\sim \lf(M_P\Lambda_b^2\over
\sigma_0\rt)^{1/4}({1/H_b}).\label{r0} \ee To have PGVSs form
after the PT, $r_0$ should satisfy $r_0< r_{*A}<1/H_b$, which
requires $\Lambda_b^2M_P< \sigma_0$ (equivalently
$H_b<H_{\sigma_0}$) and can be implemented easily. To get
(\ref{M}), we require $H_{\sigma_A}\lesssim H_b$, which is not
conflicted with $H_b<H_{\sigma_0}$, since
$\sigma_A=\sigma_0(r_0/r_{*A})^4<\sigma_0$. Given Eqs.(\ref{M})
and (\ref{r0}), we plot the dependence of ${\cal M}_{PGVS}$ on
$r_0$ for the fixed $\sigma_0=M_\sigma^3$ in Fig.\ref{fig03}.

According to (\ref{M}), ${\cal M}_{\rm PGVS}$ will be fixed by
$H_b$ (or $\Lambda_b$), if the primordial GVSs could form. If the
inflation really occurred in a Landscape with different dS vacua
\cite{Kachru:2003aw}, then the bubbles with different $\Lambda_b$
will nucleate inevitably, and one could expect that the primordial
GVSs can have a multi-peak mass spectrum. For $\Lambda_b\simeq
M_P$, we have ${\cal M}_{\rm PGVS}\sim 10^{-5}g$, which is
consistent with that of Planck-mass relic of evaporating BHs, as
proposed in Ref.\cite{MacGibbon:1987my}. However, $\Lambda_b$
should satisfy $\Lambda_b<\Lambda_{inf}\simeq
(H^2_{inf}M_P^2)^{1/4}\ll M_P$, therefore we actually have e.g.
${\cal M}_{\rm PGVS}> 10g$ for $\Lambda_{inf}\sim 10^{16}$Gev. It
is interesting that the light primordial GVSs with ${\cal M}_{\rm
PGVS}> 10^{15}g\sim 10^{-18}M_\odot$ could constitute all of cold
DM and be perfect candidates for the DM, as the GVSs do not
evaporate or evaporate much slower than BHs \cite{Raidal:2018eoo}.
The scenario proposed in this report actually provides a mechanism
responsible for such primordial GVSs.

In seminal Ref.\cite{Mazur:2001fv} proposed GVS (see also
\cite{Visser:2003ge}), it was speculated that the dS interior of
GVSs, as well as the shell with stiff fluid, could be produced by
some PT during the gravitational collapse
\cite{Nakao:2018knn}\cite{Rocha:2008hi}\cite{MartinMoruno:2011rm}.
However, we showed that the dS region of GVSs might be none other
than the interior of dS bubble nucleated during inflation. If so,
some of GVSs would be primordial, and hence could be a potential
probe to the physics of inflation.

A primordial GVS can absorb radiation and dust so that it
eventually collapses into a BH. Nonetheless, it is possible that
some of the primordial GVSs would survive till today. According to
Eq.~(\ref{M}), if $\Lambda_b\sim 0.1$Gev, we could have primordial
GVSs with ${\cal M}_{\rm PGVS}\sim 10-100M_\odot$, which could be
responsible for the LIGO/Virgo events in a similar way as
primordial BHs (see \cite{Sasaki:2018dmp} for a review on the
primordial BHs).
If such a GVS merges with a BH, we might have a post-merger GVS with a larger mass ${\cal M}>{\cal M}_{\rm PGVS}$. The addition of mass alters the shape of $V_{eff}({\tilde r}_*)$
in (\ref{V2}), so that $V_{eff}({\tilde r}_*)<0$ at ${\tilde r}_{*A}$. Assuming $\Delta {\cal M}={\cal M}-{\cal M}_{\rm PGVS}<{\cal M}_{\rm PGVS}$, at ${\tilde r}_{*A}$ we have
\ba
{d {\tilde r}_*/x\over d\tau}=-\sqrt{\lf|V_{eff}({\tilde r}_*)\rt|}&=&-\lf[\lf(1-{\tilde H}_b^2{\tilde r}_*^8\rt){\Delta{\tilde{\cal M}}\over {\tilde r}_*}+2{\tilde {\cal M}}_{\rm PGVS}{\tilde r}_*^4 \Delta{\tilde{\cal M}}\rt]^{1/2}\nonumber\\
&\simeq&-\lf(1.4{\Delta{{\cal M}}\over { {\cal M}}_{\rm PGVS}}
\rt)^{1/2}< 0, \label{velocity} \ea where we have used
Eq.(\ref{V2}) and $\Delta{\tilde{\cal M}}/ {\tilde {\cal
M}}={\Delta{\cal M}/ {\cal M}}$, as well as $V_{eff}({\tilde
r}_{*A})=0$ for ${\cal M}={\cal M}_{\rm PGVS}$. Thus the shell of
post-merger GVS (the wall of bubble) will acquire a velocity,
which is inward (or initially outward but will reverse shortly),
and the GVS will eventually collapse into a BH.

\begin{figure}[htbp]
\includegraphics[width=0.45\textwidth]{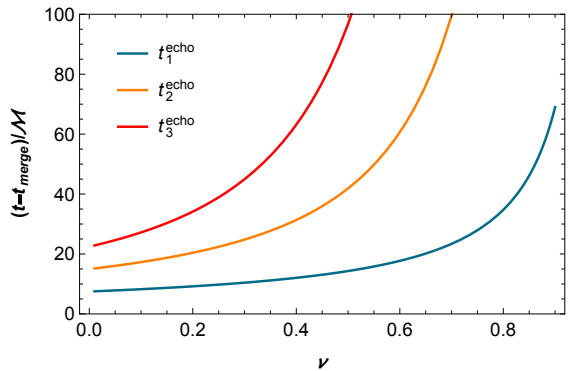}
\caption{The time at which the $n$th echo appeares approximately
with respect to the shift velocity $v$ of the shell. }
\label{fig04}
\end{figure}

It is interesting that a post-merger GVS will bring a distinct GW
echo signal. As the Schwarzschild horizon is replaced by a surface
that reflects GWs,
the reflection will result in a series of echoes in the ringdown
waveform \cite{Cardoso:2016rao}\cite{Cardoso:2016oxy}. While it is
collapsing, the post-merger GVS has the reflection surface move
towards the would-be Schwarzschild horizon. As a result, the
intervals of successive echoes will gradually increase with time
\cite{Wang:2018mlp}. Considering that when the $i$th reflection
occurred, the reflection surface has just moved to $r_i(t)$, we
can obtain that after the ringdown primary signal, the $n$th echo
will appear approximately at $t_n^{echo}-t_{merge} =2\sum_{i=1}^n
\Delta t_i$ with \be \Delta t_i \simeq
\int^{r_{i-1}}_{r_i}{dr\over v\cdot f(r)} = \int^{3{\cal
M}}_{r_i}{dr\over f(r)}, \label{Deltat} \ee where we assumed the
shift velocity $v={d{\tilde r}/ d\tau}\simeq const$ of the
reflection surface for simplicity. According to
Eq.~(\ref{Deltat}), defining $\ell_i=r_i-2{\cal M}$, we have
$\ell_i/{\cal M}\simeq (\ell_{i-1}/{\cal M})^{1\over 1-v}$
($\ell_0\simeq 0.18{\cal M}$). We plot $(t-t_{merge})/{\cal M}$
with respect to the shift velocity $v$ of the shell in
Fig.\ref{fig04}, which
suggests that at a short time-segment, if we would like to see
more than one echo, $v$ should satisfy $v\lesssim 0.95$. Together
with Eq.(\ref{velocity}), we have ${\Delta{\tilde{\cal M}}/
{\tilde {\cal M}}_{\rm PGVS}}\lesssim 0.6$. Taking the LIGO/Virgo
GW event, GW170608 $(12^{+7}_{-2}M_\odot, 7^{+2}_{-2}M_\odot)$
\cite{Abbott:2017gyy}, as a special example of primordial GVS/BH
coalescence, we have ${\Delta{{\cal M}}/ {{\cal M}}_{\rm
PGVS}}\simeq 0.6$ and $v\simeq 0.9$. Combining it with
(\ref{Deltat}), we will see that the first two echoes will appear
approximately at $t_{1,2}^{echo}-t_{merge}\simeq 5{\rm ms}, 50{\rm
ms}$, respectively. Currently, the signal-to-noise ratio of the
echo signals is too low to detect echo at enough confidence level,
however, as argued in Ref.\cite{Cardoso:2017cqb}, the proposed
Einstein Telescope has the potential of identifying such signals.


In summary, we report on a novel phenomenon that the GVSs might
come into being through the nucleation of bubbles during
inflation. The primordial GVSs not only witnessed the occurrence
of inflation, but also can survive till today and play distinct
roles in cosmological evolution. It is pointed out that if such
primordial GVSs were involved in LIGO/Virgo GW events, the
late-time GW ringdown waveform will exhibit a series of echoes,
whose intervals increasing with time. Such echo signals could be
detectable with the higher-sensitivity GWs detectors. In addition,
the inspiral stage could also be used to detect the primordial
GVSs,
e.g.\cite{Cardoso:2017cqb}\cite{Maselli:2017cmm}\cite{Pani:2010em}.

\textbf{Acknowledgments}

YSP is supported by NSFC, Nos.11575188,11690021, and also
supported by the Strategic Priority Research Program of CAS,
No.XDB23010100. YTW is supported in part by the sixty-second batch
of China Postdoctoral Fund. JZ would like to thank YSP for
hospitalities during his visit at UCAS, where this work was
initated.

 \end{document}